\documentclass[twocolumn,aps,prb,showpacs]{revtex4-1}
\usepackage{amssymb}
\usepackage{upgreek}
\usepackage[tight,TABTOPCAP]{subfigure}
\usepackage{subfloat}
\usepackage{float}
\usepackage{graphicx}
\usepackage{dcolumn}
\usepackage{bm}
\usepackage{setspace}
\usepackage{wrapfig}
\usepackage[intlimits]{empheq}
\usepackage{color}
\usepackage{relsize}

\makeatletter

\newcommand{\vect}[1]{{\boldsymbol{#1}}}
\DeclareMathOperator{\sgn}{sgn}

\begin{document}

\title{Effect of electron-phonon coupling on energy and density of states renormalizations of dynamically screened graphene}
\author{J. P. F. LeBlanc$^{1,2}$}
\email{jpfleblanc@gmail.com}
\author{J. P. Carbotte$^{3,4}$}
\author{E. J. Nicol$^{1,2}$}%
\affiliation{$^1$Department of Physics, University of Guelph,
Guelph, Ontario N1G 2W1 Canada} 
\affiliation{$^2$Guelph-Waterloo Physics Institute, University of Guelph, Guelph, Ontario N1G 2W1 Canada}
\affiliation{$^3$Department of Physics and Astronomy, McMaster
University, Hamilton, Ontario L8S 4M1 Canada}
\affiliation{$^4$The Canadian Institute for Advanced Research, Toronto, ON M5G 1Z8 Canada}
\date{\today}
\begin{abstract}
Electronic screening strongly renormalizes the linear bands which occur near the Dirac crossing in graphene.  The single bare Dirac crossing is split into two individual Dirac-like points, which are separated in energy but still at zero momentum relative to the K-point. A diamond-like structure occurs in between  as a result of the formation of plasmarons. In this work we explore the combined effect of electron-electron  and electron-phonon coupling on the renormalized energy dispersion, the spectral function and on the electronic density of states.  We find that distinct signatures of the plasmaron structure are observable in the density of states with the split Dirac points presenting themselves as minima with quadratic dependence on energy about such points.  By examining the slopes of both the density of states  and the renormalized dispersion near the Fermi level, we illustrate how one can separate $k$-dependent and $\omega$-dependent renormalizations and suggest how this might allow for the isolation of the renormalization due to the electron-phonon interaction from that of the electron-electron interaction.
\end{abstract}
%
\maketitle

\section{Introduction}
Graphene, a monolayer of carbon atoms, has been studied extensively since it was isolated in 2004.\cite{novoselov:2004}    Discoveries such as a giant Faraday rotation\cite{crassee:2011} and large strain fields which are mathematically equivalent to electrons under 300T magnetic fields\cite{levy:2010} identify graphene as a playground for studying fundamental physics.
Not least among these discoveries is the recent observation of plasmaronic peaks in the experimental angle-resolved photoemission spectroscopy (ARPES) data of Bostwick et al. \cite{bostwick:2010} for graphene epitaxially grown on H-SiC. 
Plasmarons\cite{lundqvist:1967,hedin:1967} are generally long lived quasiparticles arising from the coupling of the particle-hole continuum to charge density oscillations (plasmons).\cite{bohm:1953} They had not previously been observed in ARPES, but had been seen through other techniques in massive electron systems such as GaAs thins films and single crystal bismuth.\cite{dial:2010,tediosi:2007}
 
\begin{figure*}
  \begin{center}
\includegraphics[width=160mm]{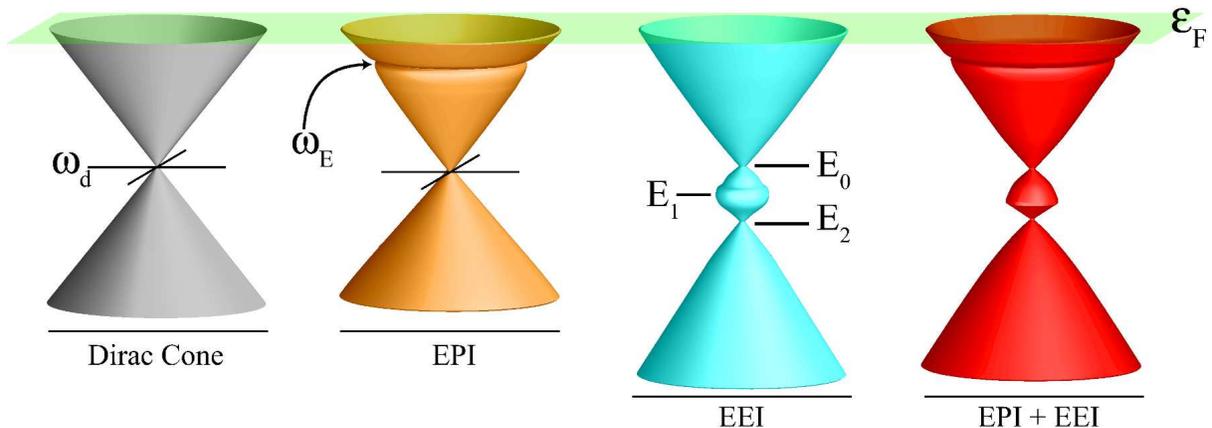}        
  \end{center}
  \caption{(Color online) Schematic illustration of the renormalizations of the Dirac cone by the electron-phonon interaction, electron-electron interaction and the combined electron-phonon and electron-electron interaction.  The EPI creates a feature at $\omega_E$. The screened EEI produces two crossing features at $E_0$ and $E_2$ with a plasmaron ring between at $E_1$.  Here, $\omega_d$ is the energy of the Dirac point and $\upvarepsilon_F$ is the Fermi level.  \label{fig:schematic}}
\end{figure*}

 In the most recent ARPES work on graphene the system is doped by the addition of potassium giving rise to  a diamond-shaped feature characterized by three energies labeled $E_0$, $E_1$, and $E_2$ (see Fig.~\ref{fig:schematic}), which, for positive doping,  we take as being implicitly negative as measured from the Fermi level.  These structures were shown to be described by calculations of electron-electron interactions (EEI) in graphene via a dynamically screened\cite{wunsch:2006} random phase approximation (G$_0$W-RPA).\cite{polini:2008,hwang:2008}  Indeed the G$_0$W-RPA results show good agreement with experiment with the exception of a disagreement at high binding energies where G$_0$W-RPA predicts two separate bands while the experiment shows these bands merging. A discussion of plasmarons in bilayer graphene is given in Ref.~\onlinecite{sensarma:2011}. 
 
Of course there are other possible interactions which could affect the electronic structure of graphene.  Perhaps the most obvious is the electron-phonon interaction (EPI) which has been calculated from first principles, and is found to be dominated by optical phonons which occur near 200 meV.\cite{park:2007}
In Fig.~\ref{fig:schematic} we illustrate schematically the primary modulations of the Dirac cone surface due to various renormalizations discussed in this paper, in order to introduce the characteristic energy scales and essential change in the Dirac cone structure. The renormalization feature of a phonon at energy $\omega_E$ and also the renormalization expected from a G$_0$W-RPA calculation of the screened Coulomb interaction are shown.  Here, rather than a simple renormalization which would modify the effective Fermi velocity in the case of the EPI, the plasmaronic side bands (not fully shown)  create a dominant diamond-shaped feature which signifies new band crossings. The inclusion of both interactions has a subtle influence on the observable plasmaronic structures.
The EPI results in $(1+\lambda)$ renormalizations, including one causing a chemical potential shift, where $\lambda$ is the electron-phonon renormalization parameter.\cite{nicol:2009,carbotte:2010}  While for an ordinary metal, this $(1+\lambda)$ factor modifies the effective mass, in graphene, the apparent result is a modification of the velocity of charge carriers, directly seen as a change in slope of the dispersion at the Fermi level.
Electron-phonon effects provide other signatures beyond $(1+\lambda)$ renormalization factors.  For coupling to an Einstein phonon of energy $\omega_E$ there are structures in the density of states and kinks (illustrated in Fig.~\ref{fig:schematic}) in the renormalized dispersion curves corresponding to this energy.\cite{nicol:2009,carbotte:2010}  There is also a background in the optical conductivity with sharp onset at $\omega_E$ due to phonon-assisted absorption, separately revealed when twice the chemical potential is larger than $\omega_E$.\cite{carbotte:2010}
 This is quite distinct from other possible effects such as band structure changes due to, for example, bilayers\cite{nicol:2008}, or from Landau quantization under an external magnetic field.\cite{gusynin:2007}  In this last case the phonon peaks in the density of states becomes further modified.\cite{pound:2011}

While it is straightforward to identify features of individual interactions it is often not clear, even to lowest order, what joint impact they have on the system.  Experimental examination of the electronic structure of graphene provides information as to the overall renormalizations but there is generally no method to disentangle the impact of different interactions.  The issue becomes what `knobs' exist in the system that can be controlled to distinguish these interactions.  
In the case of the EEI, the coupling strength (denoted by $\alpha$, analogous to the fine structure constant of quantum electrodynamics) is directly dependent on the substrate dielectric material.  Also, the self energy itself scales with doping.
  The EPI  instead has a fixed energy scale (an Einstein frequency, $\omega_E$) resulting in distinct features when the chemical potential is tuned above or below $\omega_E$.  Here, we show additional ways of extracting information on the EPI and EEI in graphene.

 In the following, we relegate to the appendix the presentation of the details of the G$_0$W-RPA theory\cite{mahan,wunsch:2006, hwangdassarma:2007, barlas:2007, polini:2008} which sums the polarization bubble to obtain the EEI self energy.  There we also present the self energy for an electron interacting with a 200 meV phonon within a simple model.\cite{park:2007}

In Section \ref{sec:AKW}, we apply these self energies and show both the renormalized energy dispersion as well as the spectral function for a variety of EEI and EPI coupling strengths.    In the absence of the phonon interaction, for strong $\alpha$ values the G$_0$W-RPA theory shows a strongly displaced band as compared to the bare band at high binding energies ($\omega<E_2$).  At energies above the crossing at $E_1$ ($\omega>E_1$) the band shows much more modest changes.  
 As $\alpha$ is reduced the strongly displaced band at high energies moves rapidly towards its non-interacting value.
Further, we find that the dispersion and spectral function are modified substantially by the addition of the phonon self energy.  In particular, the phonon coupling appears to restore weight to the bare band at high binding energies due to the fact that its self energy has the opposite sign to that of the EEI in this region which is more in agreement with what is seen in experiment.

In Section \ref{sec:DOS}, we calculate the electronic density of states (DOS) with a focus on signatures of the plasmaronic diamond structure in this quantity.  We find that the Dirac-like crossings produce parabolic features similar to those expected for a Dirac point when damping is included in the calculations.\cite{carbotte:2010,nicol:2009} 
These features shrink in amplitude for decreasing bare chemical potential, $\mu_0$, and are subject to broadening due to the EPI for $| \mu_0|>|\omega_E|$. 
 Thus, there is a narrow region of doping in which one might hope to see features of the renormalized Dirac crossings in DOS based measurements.

In Section \ref{sec:renorm}, we identify that the value of the density of states at the Fermi level is renormalized from its bare value due to the $k$-dependence of the electron-electron self energy.  The renormalized dispersion and DOS have slopes near the Fermi level with dependence on both $k$ and $\omega$ derivatives of the self energy.  We show that there is a factor difference between these two slopes which allows for the separation of $k$ and $\omega$ dependencies of these renormalizations given one has information about both the DOS and renormalized dispersion near the Fermi level.  It also allows us to separate, approximately, these two renormalization effects. It should be noted that, in this work, we restrict our discussion of renormalizations to the screened Coulomb interaction (which has both $k$ and $\omega$ dependence) and the phonon interaction (which is taken to have only $\omega$ dependence).

\section{Plasmaron features in $A(k,\omega)$ and renormalized dispersions}\label{sec:AKW}

Angle-resolved photoemission spectroscopy (ARPES) allows one to probe the charge-carrier spectral density, $A(\vect{k},\omega)$, for a given momentum, $\vect{k}$, and energy, $\omega$.
The total spectral function in graphene can be defined as the sum of two components, one for the upper band, $s=+1$, and one for the lower band, $s=-1$.  We denote these contributions as $A^+$ and $A^-$ respectively.  This results in a total spectral function  about a K-point
\begin{widetext}
\begin{equation}\label{eqn:akw}
A(k,\omega)=\sum_{s=\pm}A^s(k,\omega)=\sum_{s=\pm}\frac{1}{\pi}\frac{-{\rm Im}\Sigma_s(k,\omega) }{[\omega- {\rm Re}\Sigma_s(k,\omega)-\epsilon_k^s]^2+[{\rm Im}\Sigma_s(k,\omega)]^2}
\end{equation}\normalsize
\end{widetext}
where  $\Sigma_s(k,\omega)$ is the self energy of cone $s$ [we absorb the shift in chemical potential due to renormalization into the ${\rm Re}\Sigma_s(k,\omega)$], $\epsilon_k^s=s v_F k-\mu_0$ is the bare dispersion and $\mu_0$ is the bare chemical potential.  Note that we have taken $\hbar=1$.
Also, we have removed the explicit vector notation from the momentum, $\vect{k}$, in Eqn~(\ref{eqn:akw}).  In the cone approximation the dispersions, and therefore the spectral function, are not dependent upon the direction of $\vect{k}$, only on its magnitude, $k$.

We consider in this work two contributions to the self energy which we detail fully in the Appendix.  There we introduce the electron-electron coupling strength, $\alpha$, and the electron-phonon coupling strength, ${\rm A}$, and take the total self energy for a given band, $s$, as the sum of these two interactions.  Thus the total self energy for band $s$ is $\Sigma_s(k,\omega)=\Sigma_s^{EEI}(k,\omega)+\Sigma^{EPI}(\omega,\omega_E)$, where $\Sigma_s^{EEI}$ is the contribution of the EEI for the $s$ band which is dependent on both $k$ and $\omega$, and $\Sigma^{EPI}$ is the contribution from the EPI, which depends only on frequency $\omega$ and assumes an Einstein phonon mode at energy $\omega_E$.
The electron-electron coupling strength is given by $\alpha=\frac{g e^2}{\upvarepsilon_0 v_F}$ where $g=g_s g_v$ is the combined spin-valley degeneracy factor, $e$ is the electron charge, $\upvarepsilon_0$ is the effective dielectric constant and $v_F$ is the Fermi velocity.

As has been noted in the literature,\cite{bostwick:2010} for the case of the EEI there exists a diamond shaped feature in the renormalized dispersion which has two band energy crossings, replacing the single Dirac point of the bare case, plus a plasmaron ring.  
Clearly, in the limit of $\alpha \to 0$ (no EEI), the G$_0$W-RPA calculation must produce the bare conical dispersions. 
This is pertinent since the effective $\alpha$ goes as $\upvarepsilon_0^{-1}$, the inverse of the effective substrate dielectric constant, which is the average of that of the materials above and below the 2D graphene sheet.  In this way, the EEI in graphene becomes tunable through the substrate. 

\begin{figure}
  \begin{center}
  \includegraphics[width=85mm]{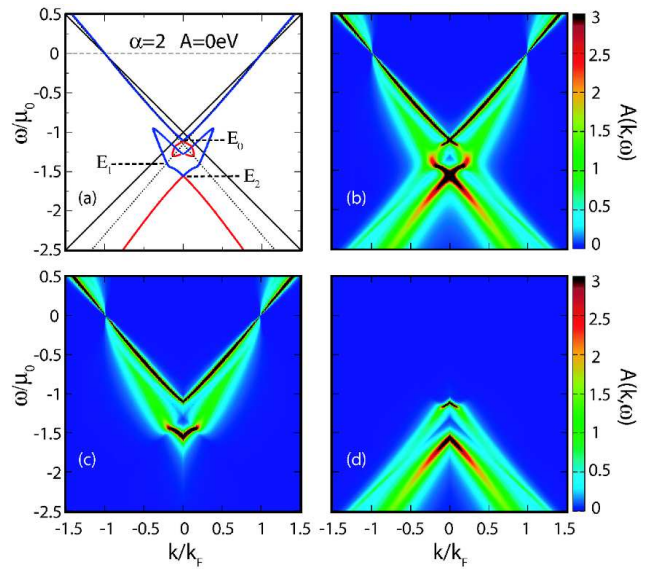}
  \end{center}
  \caption{(Color online) (a) The bare dispersion, $\epsilon_k$ (solid black lines), and EEI renormalized energies $E^+_k$ (blue) and $E_k^-$ (red) for the case of $\alpha=2$.  The upper-half cone is well approximated by a simple renormalized cone (dotted line).  (b) The total spectral density $A(k,\omega)=A^+(k,\omega)+A^-(k,\omega)$ with (c) $A^+(k,\omega)$ and (d) $A^-(k,\omega)$ shown separately. \label{fig:spectralbare}}
\end{figure}

The renormalized lines are obtained from solutions to the equation
\begin{equation}\label{eqn:poles}
\omega-{\rm Re}\Sigma_s(k,\omega)-\epsilon_k^s=0
\end{equation}
which corresponds to poles in the spectral density, $A^s(k,\omega)$, in the limit of damping going to zero.
We define the renormalized dispersion as $E_k^s=\epsilon_k^s+{\rm Re}\Sigma_s(k,\omega)$ for a given band, $s$, such that $\omega-E_k^s=0$ as in Eqn.~(\ref{eqn:poles}).
In Fig.~\ref{fig:spectralbare}(a) we compare the bare and EEI renormalized bands for the case of $\alpha=2$. 
  Here the solid blue (grey) curve applies to $s=+1$, the upper modified Dirac cone, and the red (light grey) curve to $s=-1$, the lower one.
 Note that the axes are dimensionless and scale with doping, here described through the chemical potential $\mu_0$ and the Fermi momentum, $k_F$.  
  
\begin{figure}
  \begin{center}
	\includegraphics[width=82mm]{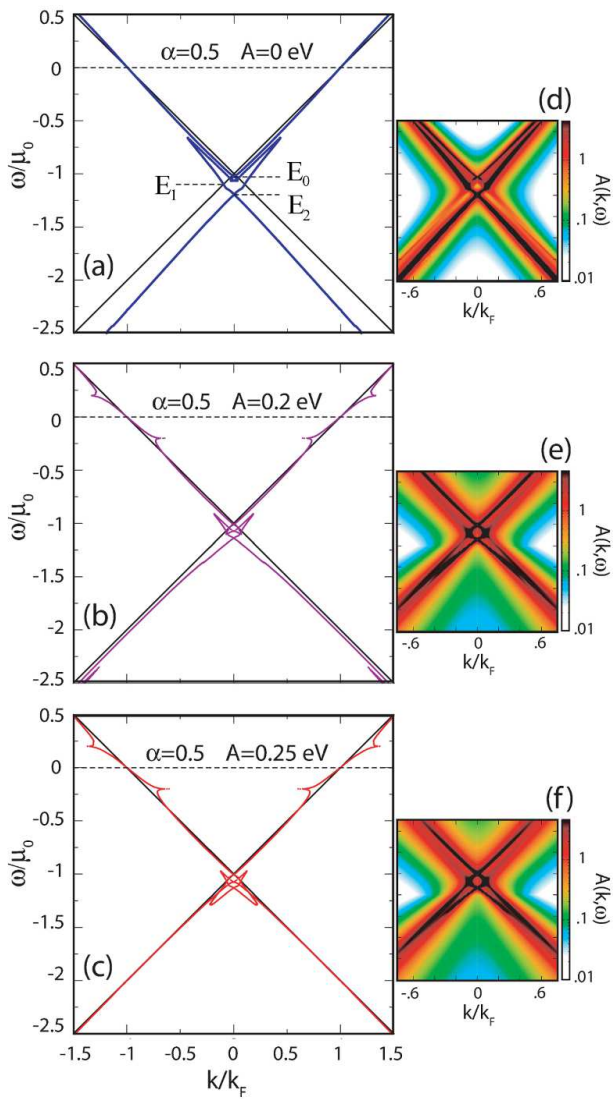}
  \end{center}
  \caption{(Color online) Energy renormalizations and spectral functions for  EEI with a fixed $\alpha=0.5$ and varied EPI coupling ${\rm A}$ for $\mu_0=1$eV.  (a-c) As the phonon strength increases, the renormalized band merges with the bare band (solid black) at high energy. The real part of the self energy due to EPI is large in this energy region and has opposite sign from that due to the EEI.  Plasmaron structures are also modified. (d-f) The spectral function for the corresponding dispersions in the region of the plasmaron feature.\label{fig:dispbare}}
\end{figure}   
  
  For most momenta, $E_k^+$ is reasonably well represented by a new cone (dotted line) with a single scaled renormalization corresponding 
  to a slightly steeper slope with the new Dirac point displaced to slightly lower energies, denoted by $E_0$.
  This is an energy regime of weak renormalizations resulting in a dispersion which is close to the bare one.
  In addition, the $k$-dependence of $\Sigma^{EEI}$ provides multiple solutions to Eqn.~(\ref{eqn:poles}) for a given $\omega$, the results of which form a complicated bat-ears-like structure in $E_k^+$ in the range $-1.5 <\omega/\mu_0 < -1.0$.  These have been identified in the literature  as representative of plasmaronic effects.\cite{bostwick:2010,polini:2008}  The bottom of this structure ends in a point which matches the top of the red curve corresponding to the lower band, $E_k^-$, together forming a second Dirac crossing at $k=0$ and energy $E_2$, below which the red curve again roughly has a conical shape, but this cone is far displaced from the bare cone for $s=-1$ (solid black curve) an indication of strong Coulomb renormalizations.  This band, $E_k^-$, also provides an additional feature which peaks at $E_0$.
  
  Fig.~\ref{fig:spectralbare}(b) shows the spectral function and Fig~\ref{fig:spectralbare}(c) and (d) its sub-components.
 The full spectral function, shown in Fig.~\ref{fig:spectralbare}(b), shows more clearly the appearance of new plasmaronic side peaks extending between $\omega=0$ and $E_2$, and then beyond but very broadened.
Primarily one notes that the Dirac point has split along $k=0$ into two separate energies at $E_0$ and $E_2$.  Comparing the renormalized dispersions of Fig.~\ref{fig:spectralbare}(a) to the color plot of Fig.~\ref{fig:spectralbare}(b) we note the familiar diamond shape.  This diamond is formed as a joint feature of the renormalized and plasmaronic bands.
Below $E_2$, the cone is strongly renormalized as this is an energy regime where transitions occur through coupling with the particle-hole continuum and plasmons.
In this energy range the spectral function again shows an additional peak not present in the renormalized energy band.
  However, comparison with the energy dispersions suggests that these peaks at larger $k$ are the residual spectral weight along the simply-renormalized Dirac cone (dotted black).

An interesting aspect of the color plots of Fig.~\ref{fig:spectralbare} is to be found in the detailed examination of the variation in the spectral intensity as a function of $k$ for fixed $\omega$. 
For frequency just above $E_0$, for instance, it is clear in Fig.~\ref{fig:spectralbare}(a) that there are three zeroes from Eqn.~(\ref{eqn:poles}) yet the intensity at these three values is quite different for each zero.
 This is a result of the $k$-dependence of the imaginary part of the self energy.  This is characteristic of the EEI and is in contrast to the EPI where the self energy has no $k$ dependence, only an energy dependence.  That the damping varies with $k$ is clearly seen in Fig.~\ref{fig:spectralbare}(b) where we note a solid black `x' shape, plus wings that are green, indicating an increase in smearing.

\begin{figure}
\begin{center}
\includegraphics[width=82mm]{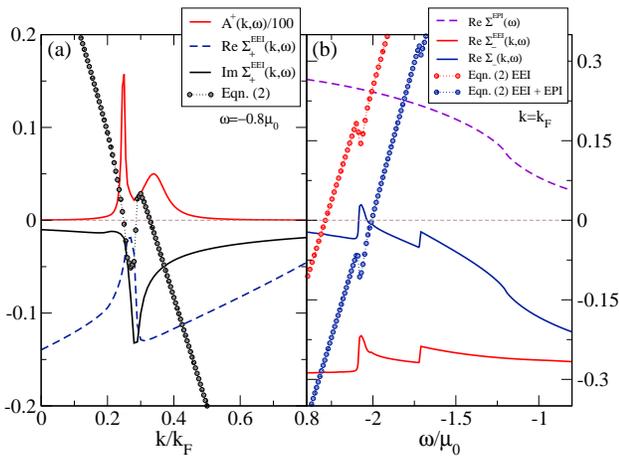}
\end{center}
\caption{(Color online) Spectral function and self energies for $\alpha=0.5$ case as in Fig.~\ref{fig:dispbare}. (a) Variation with $k$ at $\omega=-0.8\mu_0$ for $A^+(k,\omega)$ including only EEI.  The corresponding real and imaginary parts of the self energy are also shown.  The value of $\omega-{\rm Re}\Sigma -\epsilon_k$ from Eqn.~(\ref{eqn:poles}) is plotted (black circles) and shows three zeros, while the spectral function has only two obvious peaks. (b) Variation with $\omega$ at fixed $k=k_F$ of the self energies for EEI (solid red), EPI (dashed purple) and EEI+EPI (solid blue) cases for the $s=-1$ band as in Fig.~\ref{fig:dispbare}(c) and (f) as well as the values of the left side of Eqn.~(\ref{eqn:poles}) with and without the EPI (blue and red circles respectively).   \label{fig:newfig5}}
\end{figure}

\begin{figure}
\begin{center}
\includegraphics[width=82mm]{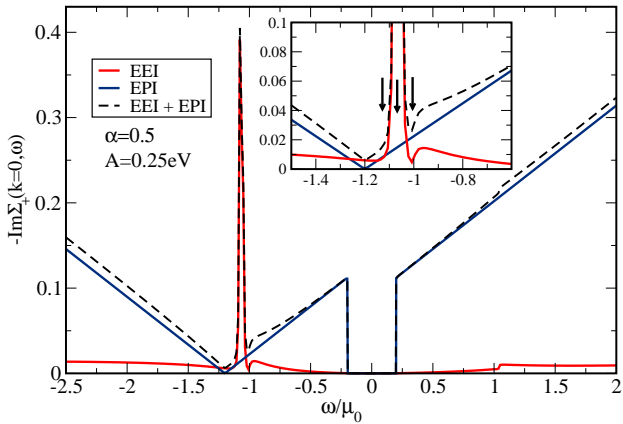}
\end{center}
\caption{(Color online) Imaginary part of the $s=1$ band self energy for $\alpha=0.5$, $A=0.25$~eV, and $\mu_0=1$eV as in Fig.~\ref{fig:dispbare}(c).  The EEI is shown in red (light grey), EPI in dark blue (dark grey) and the sum is shown as the dashed curve. Inset: Enlargement of region around $\omega/\mu_0 =-1$ with arrows indicating the locations of the three Dirac-like crossings at $k=0$ in Fig.~\ref{fig:dispbare}(c).  \label{fig:newcalc} }
\end{figure}

We now consider the case of including the EPI.  In Fig.~\ref{fig:dispbare} we display $E_k$ dispersions and $A(k,\omega)$ color map plots for the $\alpha=0.5$ case for varied EPI coupling strengths, ${\rm A}=0$, 0.2 and 0.25 [(a)$\to$(c)]. 
The black lines define the usual Dirac cone without interactions.  In Fig.~\ref{fig:dispbare}(a), by comparing the solid blue curve with Coulomb correlations for $\alpha=0.5$ with the similar curve in Fig.~\ref{fig:spectralbare}(a) we see that reducing $\alpha$ has significantly changed the shape of the plasmaron structures  of $E_k$ in the energy region around the two Dirac crossings but more importantly, below this region the bare and renormalized dispersions differ less in energy, in comparison to the $\alpha=2$ case, reflecting the reduced effect of the Coulomb interaction.
In Fig.~\ref{fig:dispbare}(b) and \ref{fig:dispbare}(c), we show how the bands are modified by the additional inclusion of the 200 meV phonon for increasing coupling strength.  We see the usual ARPES `kinks' at $\omega=\pm \omega_E$ representative of coupling to a boson.\cite{carbotte:2011,*hwang:2007,*carbotte:2005}  We can see that the plasmaronic bands have changed shape rather significantly, but most noteworthy is the now nearly unrenormalized band in the lower-half cone.  The electron-phonon interaction acts to merge these at larger binding energies. 
The renormalizations and chemical potential change of the EPI counteracts that of the EEI.  For negative values of $\omega$ the real part of the phonon self energy is positive  while the EEI contribution is negative. Thus, one expects that, for sufficiently strong phonon coupling, the joint EPI and EEI will result in energy bands nearly identical to the bare case.

This issue is further illustrated in Fig.~\ref{fig:dispbare}(d)-(f), where we display the spectral function for the cases shown in Fig.~\ref{fig:dispbare}(a)-(c) for a doping of $\mu_0=1$eV, and $\alpha=0.5$.   The diamond shaped feature is still present in the spectral function.  The height of peaks which do not appear in the renormalized dispersion will appear lighter or darker as $\omega - E_k$ gets further or closer from zero, while not being a pole, and therefore not appearing in frames (a) through (c).

 We emphasize the $k$-dependence of the EEI self energy in Fig.~\ref{fig:newfig5}(a) where we show variations with $k$ of some quantities for a given frequency $\omega=-0.8\mu_0$ which is in the region where the dispersion $E_k^+$ has additional plasmaronic peaks [as shown in Fig.~\ref{fig:dispbare}(a)].  We also include the quantity ${\omega-{\rm Re}\Sigma -\epsilon_k}$, the left hand side of Eqn~(\ref{eqn:poles}), shown as black circles.  This quantity has three zeroes at this frequency
which corresponds to three distinct lines of zeroes of Eqn.~(\ref{eqn:poles}) in the curves of Fig.~\ref{fig:dispbare}(a) on either side of $k/k_F=0$.
The zero for smallest $k$, leads to the sharp peak in $A^+(k,\omega)$ because
 the imaginary part of the self energy, shown as the solid black line in Fig.~\ref{fig:newfig5}(a) is also small.
 This is not the case for the other two zeroes, the last of which has been identified as the plasmaronic side band,  here appearing as a rather disperse peak in the spectral function near $k=0.35 k_F$.
 The second zero occurs at a point of large damping and does not show up as a peak in $A^+(k,\omega)
$.  The rapid variation of ${\rm Im}\Sigma_+^{EEI}(k,\omega)$ in this region is also the reason that the second broad peak in the spectral density (solid red curve) is shifted from the position in $k$ of the third zero.
 The partial cancellation between EEI and EPI self energy effects is detailed in Fig.~\ref{fig:newfig5}(b) where we see that in the region of interest the real parts carry opposite sign.  
 In this case the self energy depends not only on $\omega$ but also on momentum.  $\Sigma_-(k=k_F,\omega)$ is shown (red) for the EEI only case as well as the sum of EEI and EPI contributions (blue).  We see that below $\omega=-1.5$ the sum is rather small and this explains why the renormalized and unrenormalized curves approach each other.  Further, at these large binding energies there is only a single zero of Eqn.~(\ref{eqn:poles}) (solid red points) which shifts location with increasing electron-phonon coupling (solid blue points) towards positive energies.  While this is true for smaller values of $\alpha$, larger electron-electron coupling may favor multiple zeroes even at higher binding energy despite the presence of phonons.  This result of the G$_0$W-RPA calculation deviates from experimentally observed ARPES spectra across a wide range of $\alpha$ values.\cite{walter:2011}  The exact source of this deviation, or required additional interaction has yet to be understood.

The imaginary part of the self energy and how it is changed with the introduction of an electron-phonon contribution can also provide additional important information on the plasmaron structure.  In Fig.~\ref{fig:newcalc} we show results for $-{\rm Im}\Sigma_+(k=0,\omega)$ for varied $\omega$ along the $k=0$ line which defines the Dirac points.  The figure applies to the case $\alpha=0.5$ and $A=0.25$~eV which is shown in Fig.~\ref{fig:dispbare}(c) and (f).  The solid red (light grey) is the EEI contribution while the solid blue (dark grey) is the EPI contribution.  The sum of the two is represented by the dashed black curve.  Note that the EPI provides no damping in the interval from $-\omega_E$ to $\omega_E$ relative to the Fermi energy at $\omega=0$.  The physics behind this feature is that at zero temperature an electron of energy $\omega$ (assume $\omega > 0 $ for definiteness) cannot decay by the emission of a phonon if $\omega <\omega_E$.  Thus, around $\omega=0$ only electron-electron interactions contribute to the damping.  Another important feature of the EPI is that it gives, on its own, a zero in $-{\rm Im}\Sigma_+(k=0,\omega)$ at $\omega=-\omega_E -\mu_0$ which occurs at $-1.2$~eV in the case shown.  
The imaginary part is proportional to the density of states which has been shifted away from the Fermi level by $\pm \omega_E$ [see Eqn.~\ref{eqn:imsigma}] and therefore the zero in the bare quasiparticle density of states at $-\mu_0$ appears as a zero in ${\rm Im}\Sigma$ at $-\omega_E -\mu_0$.
In this region of energy, the EPI adds little to the damping which is mainly provided by the EEI which shows a strong peak just below $\omega=-\mu_0$.
This peak damps out structure in this region of the energy dispersion as seen in Fig.~\ref{fig:dispbare}(c) where there appears to be three Dirac-like crossings at $k=0$ but the middle one does not appear as a Dirac point in Fig.~\ref{fig:dispbare}(f).  This is further illustrated by the inset of Fig.~\ref{fig:newcalc} which locates with arrows the positions of the three Dirac-like crossings from Fig.~\ref{fig:dispbare}(c) and shows that the middle one lies at the energy of large damping as compared to the other two arrows which occur in a region of low damping.


\section{Plasmaron features in the Density of states}\label{sec:DOS}

The electronic density of states at a frequency $\omega$ is given in terms of the charge carrier spectral function by
\begin{align}
N(\omega)=& g_s g_{\rm v}\sum_k A(k,\omega),
\end{align}
where $g_s$ and $g_{\rm v}$ are the spin and valley degeneracies, respectively, and $A(k,\omega)$ is given by Eqn.~(\ref{eqn:akw}).  While the EPI is strongly dependent on the value of $\mu_0$, in the absence of the EPI (${\rm A}=0$), the resulting density of states including EEI is nearly independent of the value of the bare chemical potential with only a very weak dependence at large $\omega$ approaching the band cutoff, $W_c$.  In general we can write the density of states as
\begin{widetext}
\begin{align}
\frac{N(\omega)}{N_o}=& \int_{0}^{W_c} \epsilon d\epsilon \Bigg(\sum_{s=\pm}\frac{1}{\pi}\frac{-{\rm Im}{\Sigma}_s(\epsilon,\omega)}{[\omega-{\rm Re}{\Sigma}_s(\epsilon,{\omega})-s{\epsilon} +\mu_0]^2+[{\rm Im}{\Sigma}_s(\epsilon,{\omega})]^2}\Bigg), \label{eqn:dos}
\end{align}\normalsize
\end{widetext}
where \large $N_o=\frac{2}{\pi \hbar^2 v_F^2}$\normalsize.

\begin{figure}
  \begin{center}
  \subfigure{  \includegraphics[width=82mm]{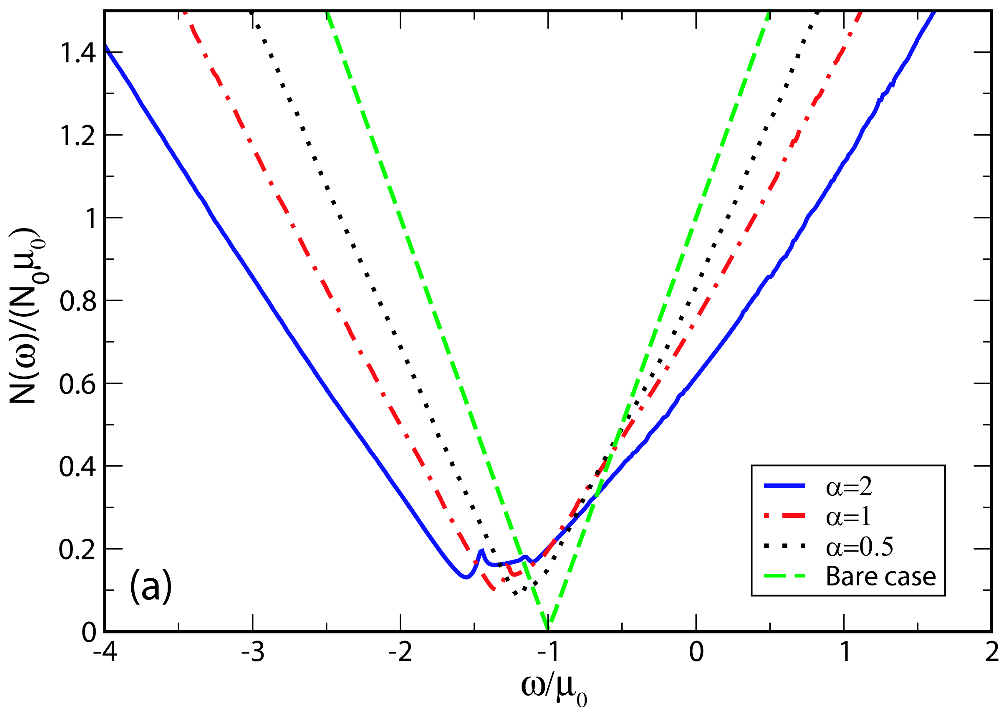}}\\
\subfigure{    \includegraphics[width=82mm]{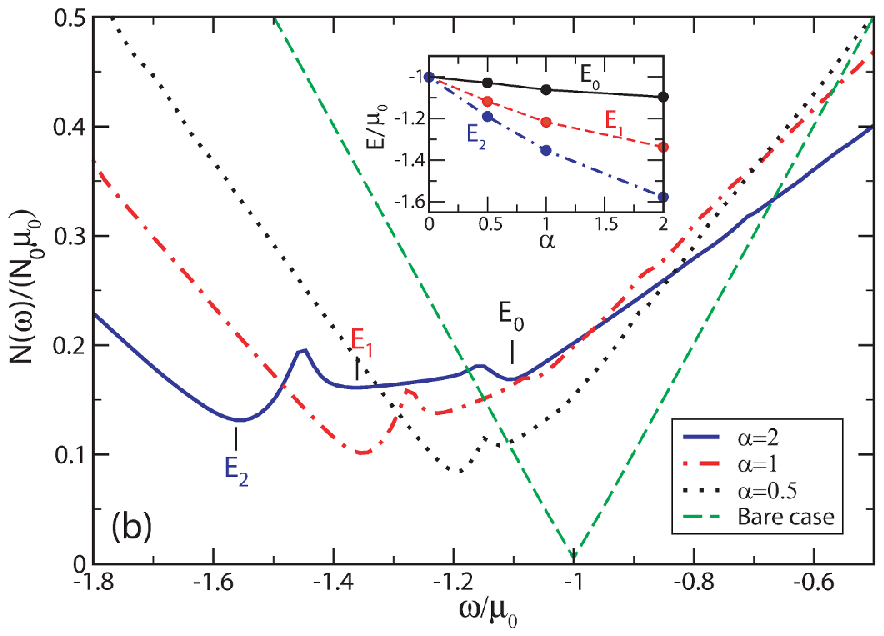}}
  \end{center}
  \caption{(Color online) DOS for EEI only. (a) Increasing $\alpha$ results in reduced slope far from the Dirac point. (b) Closeup of upper frame. Increasing $\alpha$ results in distinct parabolic features with the minima positioned at the energies corresponding to the features seen in Fig.~\ref{fig:spectralbare}(b) which are denoted as $E_0$, $E_1$ and $E_2$.  Inset:  $E_0$, $E_1$ and $E_2$  versus $\alpha$.}\label{fig:eedos}
\end{figure}

We present the results of Eqn.~(\ref{eqn:dos}) for the density of states using the G$_0$W-RPA in the absence of the EPI in Fig.~\ref{fig:eedos}, for varied  $\alpha$.  The bare DOS (green dashed line) is a wedge shape pointing to the chemical potential, with a value of zero at $\mu_0$ in the absence of scattering.  For non-zero $\alpha$, this point is lifted due to the finite lifetime at this energy provided by $|{\rm Im}\Sigma|$.\cite{carbotte:2010}  Additionally, the slope of the DOS is modified; reduced in magnitude with increasing $\alpha$ over a broad energy scale.
Furthermore, in Fig.~\ref{fig:eedos}(b) we examine the region near the Dirac point and observe distinct features in the range $\omega \in [-1.6\mu_0,-\mu_0]$. 
In the DOS, the Dirac crossings of $E_k$ at $E_0$ and $E_2$ result in a parabolic minima located at those energies as well as a similar feature at $E_1$ associated with the plasmaron ring.
It is these minima in the DOS that we use in this paper to identify the energies $E_0$, $E_1$ and $E_2$ rather than the corresponding features in the renormalized dispersion curves.
Plasmaron structures, while small in the DOS are in principle imprinted on $N(\omega)$  between the two minima associated with the split Dirac points at $k=0$.
   For variation in $\alpha$ we observe that these energies are modified.  This is plotted in the inset of Fig.~\ref{fig:eedos}(b).  The energies scale with $\mu_0$ for a given $\alpha$.  
   The energies, $E_0$, $E_1$, and $E_2$ and their variation with $\alpha$ are of importance for the experimental community, as different substrates can modify $\alpha$ through several orders of magnitude, from 2.0 for H-SiC to $10^{-3}$ for SrTiO$_3$.  In the latter case, there is evidence for strong variation of the substrate dielectric constant with temperature, creating a scenario where the value of alpha is modified by an order of magnitude as temperature is changed from room temperature to liquid helium temperature. \cite{couto:2011}

\begin{figure}
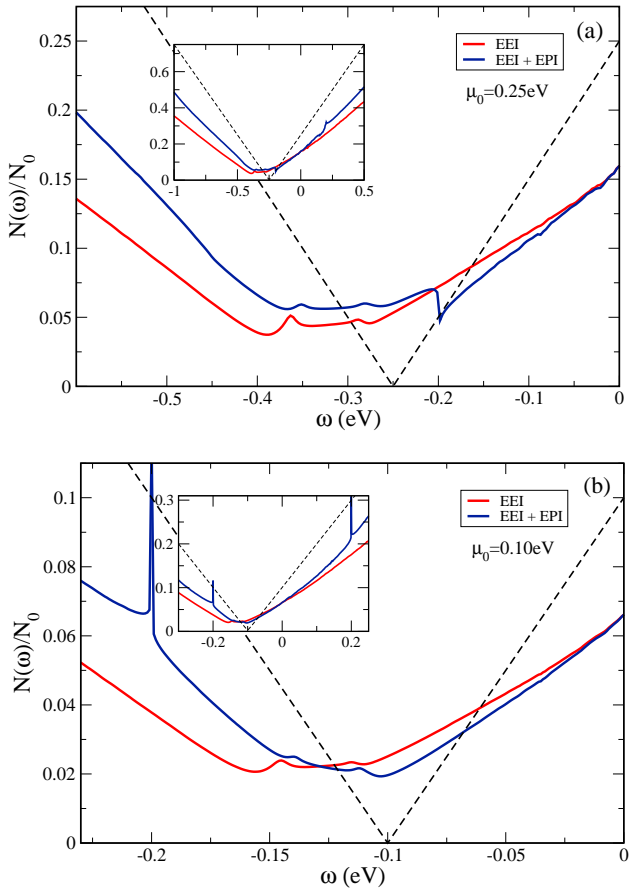

\begin{center}
\subfigure{\includegraphics[width=82mm]{./mup25-f2.eps}}
    \\
    \subfigure{ \includegraphics[width=82mm]{./mup1-f2.eps}}   
    \end{center}
    \caption{(Color online) DOS with EEI + EPI for $\alpha=2$, ${\rm A}=0.25$: (a)$\mu_0>\omega_E$, (b) $\mu_0<\omega_E$.  Phonon interaction shifts plasmaronic and Dirac features towards positive energy due to the shift in chemical potential.  Inclusion of additional phonon scattering acts to fill between these features making them less apparent in the DOS.    Insets: larger field of view of DOS.  For reference we include the bare DOS (dashed line).}\label{fig:muvary}
\end{figure}

In Fig.~\ref{fig:muvary} we display for the $\alpha=2$ case the DOS with (blue)[dark grey] and without (red)[light grey] a 200 meV phonon for chemical potential above (a) and below (b) $\omega_E$.   For too large a chemical potential, the DOS becomes smeared in the region of $E_0$, $E_1$ and $E_2$ due to additional broadening provided by the $|{\rm Im}\Sigma^{EPI}|$ resulting from electron-phonon scattering.  This is similar to the lifting of the Dirac point in the EPI only case for $\mu_0>\omega_E$.\cite{carbotte:2010}  Decreasing the chemical potential to avoid this smearing, however,  will weaken the plasmaronic features in the DOS due to the scaling with $\mu_0$.  As a result, there is perhaps a region of doping values just less than the phonon frequency where one might look for plasmaronic features in tunneling.

Several features in Fig.~\ref{fig:muvary} are to be noted.  The introduction of coupling to a phonon has shifted the structures in $N(\omega)$ corresponding to Coulomb renormalizations ($E_0$, $E_1$, $E_2$) towards positive energy though the distance between the three energies remains fairly constant.  There are also additional structures introduced which correspond to pure phonon peaks seen at $\omega=\pm\omega_E=\pm0.2$ eV.  We note also that while the slope of the DOS curve at the Fermi energy, $\omega=0$, has changed, the value of the DOS at this point is unchanged by the introduction of the EPI.  However, its EEI renormalized value is quite different from the bare density of states (dashed line).  

\begin{figure}
\begin{center}
\includegraphics[width=82mm]{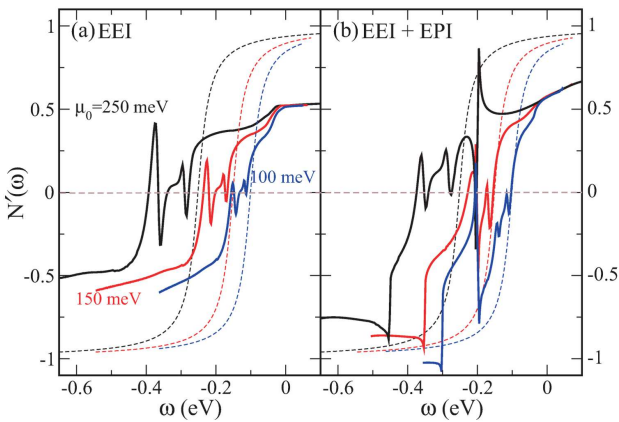}
      \end{center}
    \caption{(Color online) $N^\prime(\omega)=\frac{1}{N_0}\frac{dN(\omega)}{d\omega}$ for $\alpha=2$, ${\rm A}=0.25$ with: (a) EEI and (b) EEI +EPI, for $\mu_0=0.25$ (blue), 0.15 (red), and 0.1 eV (black).  Also shown are the bare cases (dashed lines).}\label{fig:didv}
\end{figure}

Taking a derivative with respect to $\omega$ acts to accentuate features of the parabolic minima and has been shown in the past to be an effective way to bring out subtle structures in general.\cite{nicol:2009, brar:2010}  In Fig.~\ref{fig:didv}(a), we show  $N^\prime(\omega)= \frac{1}{N_0}\frac{dN(\omega)}{d\omega}$ for three values of $\mu_0$ for the EEI with $\alpha=2$.  Also shown are the derivatives of the bare DOS, shown with dashed lines, which are smeared step functions at $\mu_0$.  In the EEI case, one immediately notes the suppressed value of $N^\prime(\omega)$ at $\omega=0$.  Further there are new features below $-\mu_0$.  These features include several zeroes and shrink in magnitude with reduced $\mu_0$
but do not change their relative shape which reflects the scaling of the self energy with $\mu_0$ noted for the pure EEI case.
Seen most clearly for $\mu_0=0.25$ eV (black), there are five zeroes below $-\mu_0$.  The zeroes which cross from negative to positive (having positive second derivative) correspond to the local minima in the DOS which are caused by the features in the spectral function at energies $E_2$, $E_1$, and $E_0$.  Thus, these zeroes describe these special energies
and represents a clear image in the quasiparticle density of states of Coulomb renormalizations and particularly of plasmarons in graphene.
If one includes the EPI as in Fig.~\ref{fig:didv}(b), there are now two new features, one occurring at $\omega=-\omega_E$ and another at $-\mu_0-\omega_E$.  This latter energy is where the $|{\rm Im}\Sigma^{EPI}|$ goes to zero. As a result of the EPI the value of $N^\prime(\omega)$ is brought closer to the bare case both at $\omega=0$ and especially so at large negative $\omega$ due to the feature at $-\mu_0-\omega_E$.  While in the EEI case the energies of the zeroes of these curves scale with $\mu_0$, in the EPI case this is not so.

\section{Renormalizations in $E_k$ and the DOS}\label{sec:renorm}
The observed impact of the EEI and EPI on slopes of the dispersion, $E_k$, and the DOS at the Fermi level as well as their variation with coupling parameters, motivates the detailed study of these renormalizations.
To do this we start by expanding the self energy about a given momenta, $k=k^*$ and energy, $\omega=\omega^*$ to lowest order. We write the real part as
\begin{align}
{\rm Re}\Sigma(k,\omega) \approx {\rm Re}\Sigma(k^*,\omega^*) +&\frac{\partial }{\partial k}{\rm Re}\Sigma(k,\omega^*)|_{k^*}(k-k^*)\nonumber \\+& \frac{\partial}{\partial \omega}{\rm Re}\Sigma(k^*,\omega)|_{\omega^*}(\omega-\omega^*).
\end{align}
We introduce the notation
\begin{align}
\lambda_{k=k^*}&=\frac{1}{v_F}\frac{\partial}{\partial k}{\rm Re}\Sigma(k,\omega^*)|_{k^*}, \\
\lambda_{\omega=\omega^*}&=-\frac{\partial}{\partial \omega}{\rm Re}\Sigma(k^*,\omega)|_{\omega^*} \label{eqn:7}.
\end{align}
From these  we define $k$ and $\omega$ dependent renormalization factors as $Z_k=1+\lambda_k$ and $Z_\omega=1+\lambda_\omega$.  At the Fermi level, Eqn.~(\ref{eqn:poles}) now becomes $\omega Z_\omega-\epsilon_kZ_k=0$.  From this we can approximate the renormalized energy in terms of the bare dispersion and the renormalization factors  as $E_k=\epsilon_k\frac{Z_k}{Z_\omega}$.

\begin{figure}
\begin{center}
\includegraphics[width=85mm]{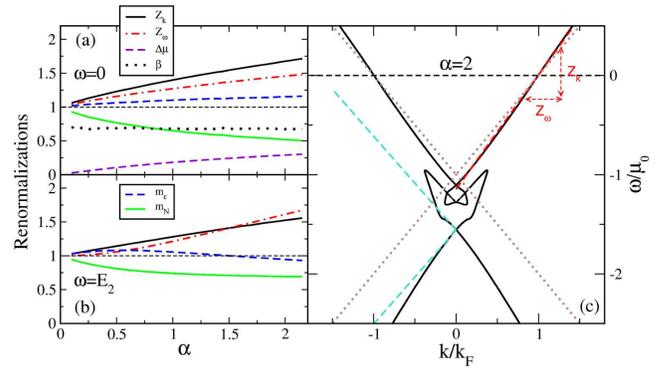}
    \end{center}
    \caption{(Color online) Renormalizations for EEI in G$_0$W-RPA for a range of $\alpha$ values shown at (a) the Fermi level and (b) $E_2$.   (c)  Energy dispersion for $\alpha=2$ (solid line) compared to the simple renormalizations at $\omega=0$ (red dashed) and $E_2$ (blue dashed).  The $\omega=0$ renormalization, shown schematically by the line with rise and run given by $Z_k$ and $Z_\omega$ respectively, roughly agrees with the full calculation until nearly $E_1$.}\label{fig:lambdas}
\end{figure}

We show details for two points on the dispersion curves, the Fermi surface ($k^*=k_F$, $\omega^*=0$) and the lower Dirac crossing ($k^*=0$, $\omega^*=E_2$).  The slopes of $E_k$ at these two points are indicated in Fig.~\ref{fig:lambdas}(c) [for $\alpha=2$] as dashed red and dashed blue, respectively.  The slope of $E_k$ at the Fermi surface defines quite well the variation of the Coulomb renormalized dispersion curves over the entire range down to the first Dirac point at $E_0$.  The dashed blue line which defines the slope of the renormalized dispersions at the second Dirac point, $E_2$,  begins to deviate somewhat from the solid blue curve for energies away from $E_2$. Thus the extent over which this approximation scheme is valid is limited near the vicinity of $E_2$.  It is worth noting that in general, the slope of this blue dashed line is not the same as that of the red dashed line.  

Fig.~\ref{fig:lambdas}(a) and (b) show slopes at the Fermi energy, $\omega=0$, and $E_2$, respectively, as a function of $\alpha$. 
  The solid (black) curve  gives $Z_k$, the dashed-dotted (red), $Z_\omega$ and the dashed (blue) $E_k/\epsilon_k$.  We define this ratio to be $E_k/\epsilon_k=m_\epsilon=Z_k/Z_\omega$, which is rather close to one for all values of $\alpha$ even though for $\alpha=2$, $Z_{k=k_F} \approx 1.7$ and $Z_{\omega=0} \approx 1.5$. 
At $\omega=0$   the $Z_\omega$ and $Z_k$ renormalizations both grow in tandem for increasing $\alpha$.  This is not the case for other energies, such as $E_2$, where above some coupling strength the renormalization values cross.  This is a consequence of the fact that the EEI energy renormalization limits to zero at the Fermi energy, not at $E_2$.  To illustrate this point, we define a quantity $\beta=\frac{\lambda^{EEI}_{\omega=0}}{\lambda^{EEI}_{k=k_F}}$ plotted as the dotted black line of Fig.~\ref{fig:lambdas}(a).  It is clear that $\beta$ is essentially constant for all values of $\alpha$ and in fact has a value of roughly $2/3$.  If one defined a similar quantity at $E_2$, or any other energy, this behavior would not be observed.  In this light, $\beta$ has a universal value for varied $\alpha$, or to put it in a more experimentally relevant context, $\beta$ is substrate independent.

   This is an important result which needs to be kept in mind when analyzing ARPES data with a view at understanding the role of Coulomb interactions.  Also shown, for completeness, is the chemical potential shift, $\Delta \mu/\mu_0$, due to the EEI.  Note that the $Z$s begin at a value of 1 for $\alpha=0$.  The lower part of the left frame gives similar results for the Dirac point at $\omega=E_2$.  
For the well known case of the electron-phonon interaction $Z_k=1$ and we recover the known result that $E_k=\epsilon_k/(1+\lambda^{\rm EPI}_\omega)$.

With this idea of slope renormalizations near the Fermi level, we draw the reader's attention back to the DOS calculations of Fig.~\ref{fig:muvary}.
We can understand both the DOS value and slope change at the Fermi level. One must rewrite the delta function $\delta(\epsilon Z_k-\omega Z_\omega)=\frac{\delta(\epsilon-\omega\frac{Z_\omega}{Z_k})}{|Z_k|}$.  Thus the density of states is subject to this $Z_k$ renormalization.

Further, in the limit $\omega \rightarrow 0$
\begin{equation}\label{eqn:nbare}
\frac{N(\omega \rightarrow 0)}{N_0}=\frac{N^{bare}(\omega \frac{Z_\omega}{Z_k})}{|Z_{k_F}|} \sim \frac{N^{bare}(0)}{|Z_{k_F}|} + \sgn(\mu_0) \frac{Z_{\omega=0}}{Z_{k_F}^2}\omega ,
\end{equation}
where $N^{bare}(\omega)=|\omega+\mu_0|$.
We conclude from Eqn.~(\ref{eqn:nbare}) that the density of states at the Fermi surface for an example case of $\alpha=2$ where $Z_{k_F}=1.63$ provides about a 40\% reduction to the DOS as seen in Fig.~\ref{fig:eedos}(a) at $\omega=0$.  This is to be contrasted to the well known effect in metal physics that the electron-phonon interaction leaves the DOS at $\omega=0$ unaltered ($Z_{k_F}^{EPI}=1$).  Further, while the variation in $N(\omega)$ for small $\omega$ remains linear in $\omega$, its slope is changed by the factor $m_N=Z_{\omega=0}/Z_{k_F}^2$ which is to be contrasted with our previous result for the dispersion curve renormalization where the renormalized quasiparticle energy, $E_k=\frac{Z_{k_F}}{Z_{\omega=0}} \epsilon_k =  m_\epsilon \epsilon_k$.   We now have slope renormalization factors for the DOS and dispersion given by $m_N$ and $m_\epsilon$, respectively.  Measuring $m_N$ in scanning tunneling spectroscopy (STS) and $m_\epsilon$ in ARPES we can separate the  renormalization factor, $ Z_{k_F}\equiv (m_\epsilon m_N)^{-1} $.  If we include both the EEI and the EPI we have that
\begin{align}
Z_k &=1+\lambda^{EEI}_k,\\
Z_\omega &= 1+\lambda^{EEI}_\omega+\lambda_\omega^{EPI}.
\end{align}
If we note that $(m_N m_\epsilon^2)^{-1}=Z_{\omega=0}= 1+\lambda^{EEI}_{\omega=0}+\lambda_{\omega=0}^{EPI}$ we can obtain  both $\lambda^{EEI}_{k_F}$ and the sum of $\lambda^{EEI}_{\omega=0}$ and $\lambda^{EPI}_{\omega=0}$ separately.  Now the value found for $\lambda^{EEI}_{k_F}$ can be used to deduce [from Fig.~\ref{fig:lambdas}(a)] the value of the coupling constant involved ($\alpha$) and consequently, the substrate dielectric constant.  Also from examination of Fig.~\ref{fig:lambdas}(a) we see that for a given $\alpha$ we can relate  $\lambda^{EEI}_{k_F}$ to $\lambda^{EEI}_{\omega=0}$ through $\beta$ such that $\lambda^{EEI}_{\omega=0}\sim \beta \lambda^{EEI}_{k_F}$.
 To continue this example, we would find that $(m_N m_\epsilon^2)^{-1} \sim 1+\beta\lambda^{EEI}_{k_F} +\lambda^{EPI}_{\omega=0}$ from which we conclude that
\begin{equation}
\lambda^{EPI}_{\omega=0}=(m_N m_\epsilon^2)^{-1}-\beta(m_N m_\epsilon)^{-1}-(1-\beta).\label{eqn:epilambda}
\end{equation}
This analysis should allow a direct experimental estimate of the strength of the electron-phonon interaction given $m_\epsilon$ measured by ARPES and $m_N$ by STS.

\begin{figure}
\begin{center}
\includegraphics[width=82mm]{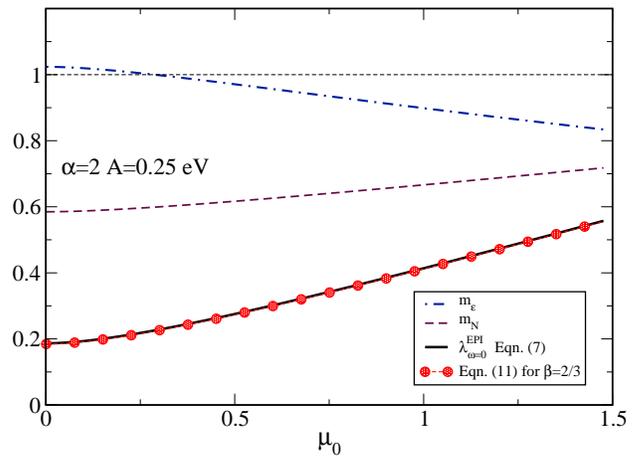}
    \end{center}
    \caption{(Color online) DOS and $E_{k_F}$ slope renormalizations, $m_N$ and $m_\epsilon$ at the Fermi level due to the EPI and EEI. Manipulation of these slopes using Eqn.~(\ref{eqn:epilambda}) allows the extraction of the EPI contribution $\lambda^{EPI}_{\omega=0}$ given knowledge of the factor $\beta$ in the EEI.  The result of Eqn.~(\ref{eqn:7}) and Eqn.~(\ref{eqn:epilambda}) match precisely due to the independence of the EEI on $\mu_0$.}\label{fig:lambda}
\end{figure}

  As we have seen, when only Coulomb effects are considered in the G$_0$W-RPA approximation there is a single curve for the density of states whatever the value of chemical potential.  As a consequence, the contribution to the renormalization at the Fermi surface from the EEI is constant for variation in $\mu_0$.    On the other hand, the electron-phonon mass renormalization increases with increasing $\mu_0$. \cite{carbotte:2010, nicol:2008}As a result, when both the EEI and EPI are considered, $m_\epsilon$ and $m_N$ will vary as seen in Fig.~\ref{fig:lambda}.  While in this theoretical work, the evaluation of Eqn.~(\ref{eqn:epilambda}) is trivial, we include the case for $\beta=2/3$ to emphasize that the knowledge of $m_\epsilon$ and $m_N$ is sufficient to obtain the EPI renormalization, $\lambda_{\omega=0}^{EPI}$, shown in Fig.~\ref{fig:lambda}. This analysis should be an experimentally realizable task, given one performs ARPES and STM on a sample with the same $\alpha$.

\section{Conclusions}
We have studied how Coulomb interactions change the density of states in graphene within the dynamically screened G$_0$W-RPA.  As was previously found for the dispersion curves, a single universal function scaled by the chemical potential $N(\omega/\mu_0)/\mu_0$ can describe all doping values for a given value of the coupling strength $\alpha$.  The Dirac point of the bare case splits into two as predicted and seen in ARPES experiments and these show up in the DOS as two minima at $E_0$ and $E_2$ with a quadratic behavior.  The plasmaronic ring presents itself as a further additional minimum at energy $E_1$ between $E_0$ and $E_2$.  These structures can serve to identify plasmarons in the DOS as measured in STS experiments.  Application of a first derivative to $N(\omega)$ further enhances these signatures.  We describe how the three energies vary with coupling $\alpha$ which is inversely proportional to the substrate dielectric material.

We find that Coulomb coupling reduces the value of the interacting DOS at the Fermi surface below its bare value by a factor of $Z_{k_F}$, independent of doping.  This is a radical departure from the well known result that the electron phonon interaction itself does not renormalize the DOS at $\omega=0$.  This difference is understood as a result of the fact that Coulomb interactions provide a self energy that depends on both momentum, $k$, and energy, $\omega$, while the EPI depends only on $\omega$.
The factor $Z_{k_F}$ comes directly from the derivative with respect to $k$ of the self energy.  On the other hand, the slope of the renormalized dispersion curves at $k_F$ is given by a ratio $Z_{k_F}/Z_{\omega=0}$, where $Z_\omega$ is related to the derivative of the self energy with respect to $\omega$.  Both factors have similar magnitudes which results in the ratio ranging from a value of 1 at $\alpha=0$ to 1.2 at $\alpha=2$.

In a similar fashion, we show that the slope of the linear-in-$\omega$ dependence of the DOS at the Fermi surface is given by the factor $Z_{\omega=0}/Z_{k_F}^2$.  Consequently a joint measurement of the slope of the dispersion curves by ARPES and that of the DOS by STS allows one to measure directly the Coulomb renormalization factor $Z_{k_F}$ as the ratio of these two slopes.

We have also studied the effect of additionally including the electron-phonon interaction.  This second interaction can significantly affect the plasmaron structure in both renormalized dispersions and density of states.  In particular, the EPI causes the energies $E_0$ and $E_2$ to move towards the Fermi level as does $E_1$ which remains roughly at the same relative position between the quadratic minima associated with the split Dirac points.  For the set of parameters examined, an important effect of including both EPI and EEI is that well below the second Dirac point, the real part of the self energy of each interaction separately carries the opposite sign, and so partially cancel against each other.  This leads to a reduction of the difference between renormalized and bare dispersion in the high binding energy region than would arise from the EEI alone.

We also establish a procedure whereby the EEI and EPI renormalizations at the Fermi surface can be separately determined from combined measurements of the DOS in STS and dispersion curves in ARPES.  


\begin{acknowledgments}
  The authors acknowledge support from the Natural Sciences and
Engineering Research Council of Canada  and the Canadian Institute
for Advanced Research.  
\end{acknowledgments}


\appendix*
\section{Theoretical Background}\label{sec:theory}
\subsection{Electron-Electron Interaction (EEI)}
One can calculate the self energy within a random phase approximation (RPA) through a procedure which is standard for a 2-dimensional electron gas (2DEG)\cite{mahan} and modified to include the issue of chirality in a chiral 2-dimensional electron gas (C2DEG).  The electron-electron interaction is assumed to be due to an effective potential, $W$, which accounts for a screened Coulomb repulsion.  The form of the self energy then includes a Green's function, $G$, and the effective screened potential, $W$, leading to the G$_0$W-RPA expression in terms of a sum over Matsubara frequencies, given by
\begin{widetext}
\begin{equation}\label{eqn:rpa}
\Sigma_s^{RPA}(\vect{k}, \omega)=-T\sum_{s^\prime=\pm, \vect{q}, m} G^o_{s^\prime}(\vect{k}+\vect{q}, \omega+\imath \Omega_m)F_{s s^\prime}(\theta_{\vect{k}\vect{k}^\prime}) W(\vect{q},\imath \Omega_m).
\end{equation}
\end{widetext}
Here the non-interacting Green's function in the $s^\prime$ band is given by
\begin{equation}
G^o_{s^\prime}(\vect{k}+\vect{q}, \omega+\imath \Omega_m)=\frac{1}{ \omega +\imath \Omega_m - \epsilon_{\vect{k}+\vect{q}}^{s^\prime}},
\end{equation}
where $\imath \Omega_m=\imath 2\pi m, m=0,\pm1,\pm2 ...$ are the bosonic Matsubara frequencies and $\epsilon_{\vect{k}^\prime}^{s^\prime}$ is the energy, relative to the Fermi level,  of the $s^\prime$ band at final momentum $\vect{k}^\prime=\vect{k}+\vect{q}$.  Due to the symmetry of the cone approximation, the $\vect{k}$ direction can define the coordinate system, such that $\epsilon_{\vect{k}^\prime}^{s^\prime}$ is a function of the magnitudes of $\vect{k}$ and $\vect{q}$ and the angle between them, $\theta_{kq}$.  This results in $\epsilon_{\vect{k}^\prime}^{s^\prime}= v_F s^\prime \sqrt{k^2+q^2+2kq \cos\theta_{kq}}-\mu_0$.  

In addition to an effective potential, the G$_0$W-RPA calculation for a chiral system includes a scattering amplitude which is the overlap of the state $k$ in the $s$ band with the state $k^\prime$ in the $s^\prime$ band.  This factor acts to remove backscattering, and is given by
\begin{equation}
F_{s s^\prime}(\theta_{\vect{k}\vect{k}^\prime})=\frac{1}{2}[1+\cos(\theta_{\vect{k}\vect{k}^\prime})s s^\prime],
\end{equation}
where $\theta_{kk^\prime}$ is the angle between the vectors $\vect{k}$ and $\vect{k^\prime}$.  Again, if one defines $k^\prime$ relative to the direction of k, we can write the angle between as
\begin{align}
\theta_{\vect{kk}^\prime}=\arctan \Bigg(\frac{\bar{q}\sin\theta_{kq}}{\bar{k}+\bar{q}\cos\theta_{kq}}\Bigg),
\end{align}
where $\bar{q}$ and $\bar{k}$ are dimensionless quantities as defined below.  We note that in calculations, it is important to correct for the non-principle value of the arctan function.

The final piece to Eqn~(\ref{eqn:rpa}) is the effective potential given by
\begin{equation}
W(q,\imath \Omega_m)=\frac{V_q}{1-V_q \Pi_0(q,\imath \Omega_m)},
\end{equation}
where $V_q$ is the 2D Coulomb potential, $V_q=\frac{2\pi e^2}{\upvarepsilon_0 q}$, where $\upvarepsilon_0$ is the effective dielectric constant of the medium and $\Pi_0$ is the polarization function for doped graphene.\cite{wunsch:2006,barlas:2007}

All necessary quantities can be written in a dimensionless form by scaling the energies by the bare chemical potential, $\mu_0$, and scaling the momenta by the bare Fermi momentum.  We denote these dimensionless quantities with a bar notation.
With this in mind we define $V_q\Pi_o= \frac{\bar{\Pi}}{\bar{q}}\cdot \alpha$, where $\alpha=\frac{ge^2}{\upvarepsilon_0 v_F}$ [$g=g_sg_{\rm v}$ is the combined spin and valley degeneracy factor] and $\bar{\Pi}$ is given for a given $\bar{q}$ and $\imath \bar{\Omega}$ as \cite{barlas:2007}
\begin{widetext}
\begin{align}
\bar{\Pi}&(\bar{q},\imath \bar{\Omega})= -1-\frac{\pi}{8}\frac{\bar{q}^2}{\sqrt{\bar{\Omega}^2+\bar{q}^2}}\nonumber\\&+\frac{1}{4}\frac{\bar{q}^2}{\sqrt{\bar{\Omega}^2+\bar{q}^2}}  {\rm Re} \left[\arcsin\left(\frac{2+\imath\bar{\Omega}}{\bar{q}}\right)+\left(\frac{2+\imath\bar{\Omega}}{\bar{q}}\right)\sqrt{1-\left(\frac{2+\imath\bar{\Omega}}{\bar{q}}\right)^2}\right]\label{eqn:polnorm}.
\end{align}
\end{widetext}
\normalsize
Alternate polarization functions have been explored such as, for example, the case of gapped graphene\cite{qaiumzadeh:2009} which maps onto Eqn.~(\ref{eqn:polnorm}) as the gap goes to zero.
\noindent In this case we can write the effective potential as
\begin{equation}
W(\bar{q},\imath \bar{\Omega})=\frac{2\pi \alpha}{g}\frac{v_F}{k_F}\frac{1}{\bar{q}-\alpha \bar{\Pi}(\bar{q},\imath \bar{\Omega})}.
\end{equation}

Evaluation of the self energy of Eqn.~(\ref{eqn:rpa}) is generally performed in two parts, identified as the line and residue (RES) portions.\cite{mahan}  The line component is completely real while the residue portion has both real and imaginary parts. These are given by
\begin{widetext}
\begin{align}\label{eqn:res}
\bar{\Sigma}_s^{RES}(\bar{k},\bar{\omega})=\sum_{s^\prime=\pm 1}\int_{0}^\infty \int_{0}^{2\pi} &\frac{d\bar{q}d\theta_{kq}}{2\pi}\frac{\alpha}{g}\upvarepsilon^{-1}(\bar{q},\bar{\omega}-\bar{\epsilon}_{\vect{k}+\vect{q}}^{s^\prime})F_{ss^\prime}(\theta_{\vect{kk}^\prime})\times\nonumber \\&[\Theta(\bar{\omega}-\bar{\epsilon}_{\vect{k}+\vect{q}}^{s^\prime})-\Theta(-\bar{\epsilon}_{\vect{k}+\vect{q}}^{s^\prime})],
\end{align}
and 
\begin{align}\label{eqn:line}
\bar{\Sigma}_s^{line}(\bar{k},\bar{\omega})=&-\sum_{s^\prime=\pm 1}\int_{0}^\infty \int_{0}^{2\pi}\frac{d\bar{q} d\theta_{kq}}{2\pi}\frac{\alpha}{g}F_{ss^\prime}(\theta_{\vect{kk}^\prime})\times \nonumber\\ &\int_{-\infty}^{\infty}\frac{d\bar{\Omega}}{2\pi}\upvarepsilon^{-1}(q,\imath \bar{\Omega})\Bigg[\frac{\bar{\omega}-\bar{\epsilon}_{\vect{k}+\vect{q}}^{s^\prime}}{\bar{\Omega}^2+(\bar{\epsilon}_{\vect{k}+\vect{q}}^{s^\prime}-\bar{\omega})^2}-\frac{\imath \bar{\Omega}}{\bar{\Omega}^2+(\bar{\epsilon}_{\vect{k}+\vect{q}}^{s^\prime}-\bar{\omega})^2}\Bigg],
\end{align}
\end{widetext}
\normalsize
where $\upvarepsilon^{-1}$ is given by
\begin{equation}
\upvarepsilon^{-1}=\frac{1}{1-V_q\Pi_o}=\frac{\bar{q}}{\bar{q}-\alpha \bar{\Pi}}\label{eq:die}.
\end{equation}
Here we have also written the band energies for general $\vect{k}$ and $\vect{q}$ in a dimensionless form
\begin{align}
\bar{\epsilon}_{\vect{k}+\vect{q}}^{s^\prime}=\frac{\epsilon_{\vect{k}+\vect{q}}^{s^\prime}}{\mu_0}=s^\prime\sqrt{\bar{k}^2+\bar{q}^2+2\bar{k}\bar{q}\cos\theta_{kq}}-1
\end{align}
\normalsize
which is the energy in the Dirac cone approximation. 

The integral over $\bar{\Omega}$ in Eqn.~(\ref{eqn:line}) contains two terms,  the first has a Lorentzian shape which provides a natural cutoff while the second term is odd and vanishes over the $\bar{\Omega}$ integral.  
Also, $\bar{\Pi}$ for purely imaginary frequencies is completely real.\cite{barlas:2007}  The evaluation of real frequencies in, for example, $\upvarepsilon^{-1}(\bar{q},\bar{\omega})$ requires that one include a finite lifetime for the plasmons.  This can be done in the form $\upvarepsilon^{-1}(\bar{q},\bar{\omega}) \to \upvarepsilon^{-1}(\bar{q},\bar{\omega} +\imath \Gamma)$, where $\Gamma$ is the plasmon scattering rate.\cite{wunsch:2006,pyatkovskiy:2009} Failure to include such a term results in infinitely long lived plasmons which cannot scatter with electrons.  Thus, such a rate controls electron-plasmon coupling (plasmarons).

Previous work\cite{hwang:2008} has shown that the $q$-integrals of Eqn.~(\ref{eqn:res}) and (\ref{eqn:line}) suffer from a divergence at large $q$.  This ultraviolet divergence is an artifact of the assumed linear dispersion in graphene, which allows for arbitrarily large $q$ scattering.  A detailed calculation would include a band cutoff at an energy $W_c\simeq7$eV which corresponds to the energy which maintains the number of states in the Brillouin zone.  Hence, this is important for density of states calculations, such as Eqn.~(\ref{eqn:dos}) where we include a normalized ultraviolet cutoff of $\Lambda=\pm W_c/\mu_0$. A similar cutoff is required in the self energy calculation itself.  It has been established\cite{polini:2007} that the value of $\Lambda$ lies in the range of $10\to 100$ for experimentally relevant values of chemical potential. Thus we have chosen a fixed cutoff of $\Lambda=50$ in the self energy formulae, independent of the value of  $\mu_0$.  This fixed cutoff removes any $\mu_0$ dependence in the EEI self energy. If one seeks to compare directly to experiment, then $\Lambda$ can be modified for a given doping.  The precise variation due to the choice of $\Lambda$ was thoroughly explored in Ref.~\onlinecite{polini:2007}.
 

We define the total self energy for band $s$ due to the electron-electron interaction as
\begin{equation}
\bar{\Sigma}_s^{EEI}(\bar{k},\bar{\omega}) = \bar{\Sigma}_s^{RES}(\bar{k},\bar{\omega}) +\bar{\Sigma}_s^{line}(\bar{k},\bar{\omega}).
\end{equation}
One can reinstate units of energy by writing $\Sigma_s^{EEI}(k,\omega)=\mu_0 \bar{\Sigma}_s^{EEI}(\bar{k},\bar{\omega})$.

\subsection{Electron-Phonon Interaction (EPI)}

We take the total self energy to be the sum of the EEI and EPI contributions such that the total is $\Sigma_s=\Sigma^{EEI}_s+\Sigma^{EPI}-\Delta\mu$, where $\Delta\mu$ is the correction to the chemical potential (given by the real part of the self energies evaluated at $k=k_F$ and $\omega=0$).  While the electron-electron contribution is dependent on $k$, the electron-phonon interaction is not.  The EPI self energy is given by\cite{carbotte:2010}

\begin{widetext}
\begin{eqnarray}\label{eqn:resigma}
&{\rm Re}\Sigma_{\mu_0>0}^{EPI}(\omega ,\omega_E)=\displaystyle
\frac{{\rm A}}{W_c}\Biggl\{\omega_E \ln\Biggl|
\displaystyle
\frac{(W_c+\omega_E-\omega-\mu_0)(\mu_0+\omega+\omega_E)^2}{(\omega^2-\omega_E^2)
(W_c+\omega+\omega_E+\mu_0)}\Biggr|\nonumber\\
&-(\mu_0+\omega)\ln\Biggl|
\displaystyle
\frac{(W_c+\omega_E-\omega-\mu_0)(W_c+\omega+\omega_E+\mu_0)(\omega+\omega_E)}
{(\omega-\omega_E)(\omega+\mu_0+\omega_E)^2}\biggr|\Biggr\}
\end{eqnarray}
\end{widetext}
for the real part, where
\begin{equation}
{\rm Re}\Sigma^{EPI}_{\mu_0<0}(\omega ,\omega_E)=-{\rm Re}\Sigma^{EPI}_{|\mu_0|}(-\omega ,\omega_E).
\end{equation}
The imaginary part is given by 
\begin{widetext}
\begin{equation}\label{eqn:imsigma}
-{\rm Im}\Sigma^{EPI}(\omega ,\omega_E)=
\begin{cases}\displaystyle\frac{\pi {\rm A}}{W_c}|\omega-\omega_E+\mu_0|, & {\rm for}\quad \omega_E<\omega<W_c-\mu_0+\omega_E ,\\
           \displaystyle\frac{\pi {\rm A}}{W_c}|\omega+\omega_E+\mu_0|, & {\rm for}\quad -\omega_E>\omega>-W_c-\mu_0-\omega_E ,
\end{cases}
\end{equation}
\end{widetext}
and is zero outside these intervals.   We see that this self energy is a function of frequency, $\omega$, and also dependent upon the choice of Einstein frequency, $\omega_E$.
 We assume a model where $\omega_E=200$ meV as has been done previously \cite{carbotte:2010} which was proposed by Park et al\cite{park:2007} on the grounds of fitting density functional theory calculations within the local density approximation (LDA).

\bibliographystyle{apsrev4-1}
\bibliography{bib}

\end{document}